\documentclass[twocolumn,english]{revtex4-2}
\usepackage[T1]{fontenc}
\usepackage[latin9]{luainputenc}
\usepackage{amsmath}
\usepackage{graphicx}
\usepackage{babel}
\begin{document}
\title{Time reversal invariant topological 1D and 2D superconductors: doubling
the Sau-Luchtin-Tewari-Sarma \citep{Sau_2009} and Oreg-Refael-von
Oppen \citep{Oreg_2010,Luchtyn_2010} proposals}
\author{Garry Goldstein$^{1}$}
\address{$^{1}$garrygoldsteinwinnipeg@gmail.com}
\begin{abstract}
In this work we present doubled versions of the Sau-Luchtin-Tewari-Sarma
\citep{Sau_2009} and Oreg-Refael-von Oppen \citep{Oreg_2010,Luchtyn_2010}
proposals thereby obtaining time reversal invariant p-wave superconductivity
in both 1D and 2D. This construction is much like the Kane-Mele spin
Hall model \citep{Kane_2005}, which is a time reversal invariant
doubling of the Haldane model \citep{Haldane_1988}. We show that
the low energy effective action for these doubled versions of the
Sau-Luchtin-Tewari-Sarma \citep{Sau_2009} and Oreg-Refael-von Oppen
\citep{Oreg_2010,Luchtyn_2010} models correspond to single band p-wave
time reversal invariant superconductors with pseudospin degree of
freedom instead of spin degree of freedom. There are Majorana fermions
at the ends of wires or in vortex cores of these superconductors.
Furthermore these Majorana fermions are shown to be stable to small
perturbations. In the supplement we present a physical realization
of the system with cold atoms \citep{Jiang_2011} and show that a
related ``no-go'' theorem given in Ref. \citep{Haim_2019,Haim_2016}
has too restrictive assumptions to apply to this proposal.
\end{abstract}
\maketitle

\section{\protect\label{sec:Introduction}Introduction}

The scientific community has been studying topological superconductivity
and superfluidity for a relatively long time. It is now known that
the $^{3}He$ -A and -B phases are topological superfluids. The $^{3}He$
-A and -B phases have been characterized by topological bulk invariants
\citep{Volovik_2003}. $p+ip$ superconductors in 2D also posses two
topologically invariant phases and exhibit Majorana fermions on edges
between them \citep{Read_2000}. The smoking gun for topological superconductors
and superfluids is the existence of zero energy Majorana modes in
the vortices of their order parameters \citep{Alicea_2010,Bernevig_2013,Haim_2016,Keselman_2013,Lindler_2010,Luchtyn_2010,Oreg_2010,Qi_2010,Sarma_2021,Sau_2009,Shen_2017,Volovik_2003,Pan_2020}.
Majorana fermions are real fermions which are their own antiparticles
\citep{Majorana_1937}. Sau et. al. \citep{Sau_2009} suggested creating
Majorana fermions in vortices of ferromagnetic insulator/semiconductor/s-wave
superconductor superstructures. The authors of \citep{Sau_2009} showed
that Majorana fermions exist in this setup by solving the single vortex
core problem for the effective Hamiltonian for the superstructure
\citep{Sau_2009}. Alicea \citep{Alicea_2010} extended the work of
Sau et al. \citep{Sau_2009} to replace the external magnetic field
with a ferromagnet. Shen presented \citep{Shen_2017} an equivalence
between the model in Sau et. al. \citep{Sau_2009} and spinless $p+ip$
superconductors. In this work for clarity in the supplement we extend
these results and show that the low energy effective action for large
magnetic fields is that of a spinless $p+ip$ superconductor \citep{Shen_2017}
(however this is not the main thrust of this work). These ideas about
2D s-wave superconductor proximitized semiconductors were extended
to 1D by \citep{Oreg_2010,Luchtyn_2010} where it was shown that such
ferromagnetic insulator/semiconductor/s-wave superconductor heterostructures
but with 1D semiconductor wires have Majorana modes at the ends of
the wires much like 1D spinless p-wave superconductors. In this work,
in the supplement \citep{Supplement_2024}, we show that the low energy
effective action is that of a 1D spinless p-wave superconductor as
well. 

The main thrust of this this work is to extend the ideas of Sau et
al. \citep{Sau_2009} as well as the 1D case of Refs. \citep{Oreg_2010,Luchtyn_2010}
to the case of two band s-wave superconductor proximitized semiconductors
with both Kane Mele like \citep{Kane_2005} and Rashba \citep{Manchon_2015}
spin orbit coupling. Furthermore no large magnetic fields (or ferromagnets)
are involved as in the Sau-Luchtin-Tewari-Sarma \citep{Sau_2009}
and Oreg-Refael-von Oppen \citep{Oreg_2010,Luchtyn_2010} proposals
- which is highly advantageous to superconductivity \citep{Coleman_2016}.
We replace the time reversal symmetry breaking effects of the magnetic
field with spin orbit coupling (which is time reversal (TR) invariant)
thereby doubling the proposals in Refs. \citep{Sau_2009,Oreg_2010,Luchtyn_2010}
in a TR invariant way. This is much like the Kane-Mele spin Hall proposal
\citep{Kane_2005} doubled the Haldane model \citep{Haldane_1988}
in a TR invariant way. We show that to leading order the low energy
effective action in 2D is time reversal invariant p-wave superconductivity
with two copies of $p+ip$ superconductors with opposite chirality,
the two copies are in pseudospin space rather then in spin space as
is the usual case with TR invariant superconductors \citep{Shen_2017,Bernevig_2013}.
Furthermore in the supplement we show that ``no-go'' theorems derived
previously \citep{Haim_2019,Haim_2016} make assumptions that are
not general enough to cover proximity effects from multi-band superconductors
considered here and as such do not apply to the setup presented in
this work. Majorana fermions are found in vortex cores for the superconductors
in this work and their stability is proved for small perturbations.
In 1D, for the superconductors in this work, we find Majorana fermions
at the end of wires (these are also stable to small perturbations)
with the effective low energy theory of the doubled Oreg-Refael-von
Oppen \citep{Oreg_2010,Luchtyn_2010} proposal is that of a single
band time reversal invariant p-wave superconductor with pseudospin
rather then spin degree of freedom.

\section{\protect\label{sec:General-considerations-about}General considerations
about time reversal and the type of orbitals needed for our construction}

In order to obtain the doubled Sau-Luchtin-Tewari-Sarma \citep{Sau_2009}
and Oreg-Refael-von Oppen \citep{Oreg_2010,Luchtyn_2010} proposals
we require two spinful orbitals with specific time reversal properties
given by Eq. (\ref{eq:Time_reversal_transform}) below. Any two orbitals
with this transform property under time reversal will do. Here we
will show that these time reversal properties are not exotic by presenting
an explicit physical realization of such orbitals with $\left|Y_{1}^{1}\right\rangle ,\,\left|Y_{-1}^{1}\right\rangle $
(which transform under TR as in Eq. (\ref{eq:Time_reversal_transform})
). Here $Y_{m}^{l}$ are spherical harmonics. For the rest of the
paper, for simplicity, the reader may focus on this realization if
they so please, though many others are possible.

\subsection{\protect\label{subsec:Physical-setup}Physical setup}

We now present explicit orbitals with time reversal properties given
by Eq. (\ref{eq:Time_reversal_transform}) below. For concreteness
we will consider the case of $p_{x}$ and $p_{y}$ orbitals which
are a linear combination of $\left|Y_{1}^{1}\right\rangle ,\,\left|Y_{-1}^{1}\right\rangle $
. We will work in the basis: 
\begin{equation}
\left|+\right\rangle =\left|Y_{1}^{1}\right\rangle \sim\exp\left(i\theta\right),\;\left|-\right\rangle =\left|Y_{-1}^{1}\right\rangle \sim-\exp\left(-i\theta\right).\label{eq:Basis}
\end{equation}
Here $\theta$ is the azimuthal angle. For each lattice site for the
systems we will consider there will be four relevant basis states:
\begin{equation}
\left|+,\uparrow\right\rangle ,\:\left|+,\downarrow\right\rangle ,\:\left|-,\uparrow\right\rangle ,\:\left|-,\downarrow\right\rangle \label{eq:Basis_spinful}
\end{equation}
We will use $\tau$ as the Pauli matrices within the orbital space
and $\sigma$ to be the Pauli matrices within the spin space, below
we shall also introduce $\mu$ which are Pauli matrices for particle-hole
space. We note that $\left|Y_{1}^{2}\right\rangle ,\;\left|Y_{-1}^{2}\right\rangle $
are just as good for our purposes of obtaining Eq. (\ref{eq:Time_reversal_transform})
as well as many other orbitals.

\subsection{\protect\label{subsec:Time-reversal-symmetry-2}Time reversal symmetry}

Under time reversal symmetry we have that $\vec{\sigma}\rightarrow-\vec{\sigma}$.
Furthermore under time reversal $Y_{m}^{l}\rightarrow\left(-1\right)^{m}Y_{-m}^{l}$
(as $i\rightarrow-i$, see Eq. (\ref{eq:Basis})) which means that
we have that in the model we propose below the time reversal operator
is given by 
\begin{equation}
T:-\tau_{x}i\sigma_{y}K\label{eq:Time_reversal}
\end{equation}
Where $K$ is complex conjugation. This means that we have 
\begin{equation}
T:\tau_{z}\rightarrow-\tau_{z},\,\tau_{x}\rightarrow\tau_{x},\,\tau_{y}\rightarrow\tau_{y}\label{eq:Time_reversal-1}
\end{equation}
Furthermore under time reversal $\mathbf{k}\rightarrow-\mathbf{k}$
(here $\mathbf{k}$ is the pseudo-momentum in the first Brillouin
zone). As such we have that: 
\begin{align}
T: & c_{\mathbf{k}+\uparrow}^{\dagger}\rightarrow-c_{-\mathbf{k}-\downarrow}^{\dagger},\:c_{\mathbf{k}+\downarrow}^{\dagger}\rightarrow+c_{-\mathbf{k}-\uparrow}^{\dagger},\nonumber \\
 & c_{\mathbf{k}-\uparrow}^{\dagger}\rightarrow-c_{-\mathbf{k}+\downarrow}^{\dagger},\:c_{\mathbf{k}-\downarrow}^{\dagger}\rightarrow+c_{-\mathbf{k}+\uparrow}^{\dagger}\label{eq:Time_reversal_transform}
\end{align}

\section{\protect\label{sec:Doubling-the-Sau-Luchtin-Tewari-}Doubling the
Sau-Luchtin-Tewari-Sarma proposal (p-wave time reversal preserving
Hamiltonian)}

\subsection{\protect\label{subsec:Global-analysis}Global analysis}

We consider a two band spinful BDG Hamiltonian given by: 
\begin{align}
H\left(\mathbf{k}\right) & =\mu_{z}\left[\left(\frac{\mathbf{k}^{2}}{2m}-\mu\right)+\lambda_{SO}\sigma_{z}\tau_{z}+\lambda_{R}\left[\mathbf{k}_{y}\sigma_{x}-\mathbf{k}_{x}\sigma_{y}\right]\right]\nonumber \\
 & +\Delta_{s}\mu_{x}\label{eq:Hamiltonian-3-2}
\end{align}
We note that $\left[H\left(\mathbf{k}\right),\tau_{z}\right]=0$ so
the two orbitals decouple and the Hamiltonian is doubled and $\lambda_{SO}\sigma_{z}\tau_{z}$
is similar to the spin orbit coupling used by Kane and Mele \citet{Kane_2005}
except $\tau_{z}$ is not in the valley space for graphene \citet{Kane_2005}.
We now use the basis: 
\begin{align}
\Psi_{\mathbf{k}}^{\tau_{z}}= & \left(c_{\mathbf{k}+\uparrow},c_{\mathbf{k}+\downarrow},c_{-\mathbf{k}+\downarrow}^{\dagger},-c_{-\mathbf{k}+\uparrow}^{\dagger}\right.\nonumber \\
 & \left.c_{\mathbf{k}-\uparrow},c_{\mathbf{k}-\downarrow},c_{-\mathbf{k}-\downarrow}^{\dagger},-c_{-\mathbf{k}-\uparrow}^{\dagger}\right)^{T}\label{eq:Basis_decoupled}
\end{align}
As such:

\begin{equation}
H\left(\mathbf{k}\right)=\left(\begin{array}{cc}
H_{+}\left(\mathbf{k}\right) & 0\\
0 & H_{-}\left(\mathbf{k}\right)
\end{array}\right)\label{eq:Block_diagonal}
\end{equation}
with 
\begin{align}
H_{\pm}\left(\mathbf{k}\right)= & \mu_{z}\left[\left(\frac{\mathbf{k}^{2}}{2m}-\mu\right)\pm\lambda_{SO}\sigma_{z}+\lambda_{R}\left[\mathbf{k}_{y}\sigma_{x}-\mathbf{k}_{x}\sigma_{y}\right]\right]\nonumber \\
 & +\Delta_{s}\mu_{x}\label{eq:Two_copies}
\end{align}
being two time reversed copies of the Sau-Luchtin-Tewari-Sarma Hamiltonians
(the total Hamiltonian is TR invariant see Eq. (\ref{eq:Time_reversal_transform})).
Furthermore since this is an exact doubling the phase boundaries (where
the Hamiltonian become gapless) are identical to those in the proposal
by Sau et. al. \citep{Sau_2009} with the critical boundary given
by \citep{Sau_2009}: 
\begin{equation}
\mu=\sqrt{\lambda_{SO}^{2}-\Delta_{s}^{2}}\label{eq:Critical}
\end{equation}

\subsection{\protect\label{subsec:Effective-low-energy}Effective low energy
theory}

\subsubsection{\protect\label{subsec:Hamiltonian-no-pairing}Hamiltonian ignoring
the effects of pairing}

We now introduce the spinor $\Psi_{\mathbf{k}}=\left(c_{\mathbf{k}+\uparrow},c_{\mathbf{k}+\downarrow},c_{\mathbf{k}-\uparrow},c_{\mathbf{k}-\downarrow}\right)^{T}$
then we consider the Hamiltonian given by: 
\begin{equation}
H\left(\mathbf{k}\right)=\Psi_{\mathbf{k}}^{\dagger}\mathcal{H}\left(\mathbf{k}\right)\Psi_{\mathbf{k}}\label{eq:Hamiltonian_matrix-2}
\end{equation}
with:
\begin{equation}
\mathcal{H}\left(\mathbf{k}\right)=\left(\frac{\mathbf{k}^{2}}{2m}-\mu\right)+\lambda_{SO}\tau_{z}\sigma_{z}+\lambda_{R}\left(\mathbf{k}_{x}\sigma_{y}-\mathbf{k}_{y}\sigma_{x}\right)\label{eq:4_by_4}
\end{equation}
It is straightforward to check that $\left[\mathcal{H}\left(\mathbf{k}\right),\tau_{z}\right]=0$
and that the Hamiltonian in Eq. (\ref{eq:4_by_4}) is TR invariant
and corresponds to the non-pairing piece of the Hamiltonian in Eq.
(\ref{eq:Hamiltonian-3-2}). Now we will assume that $\lambda_{SO}\gg\lambda_{R}$
so that the low energy subspace of the Hamiltonian has a basis given
by \citep{Supplement_2024}: 
\begin{align}
\left|\Uparrow\right\rangle  & =\left|+,\downarrow\right\rangle -\frac{\lambda_{R}}{2\lambda_{SO}}\left(i\mathbf{k}_{x}-\mathbf{k}_{y}\right)\left|+,\uparrow\right\rangle +...\nonumber \\
\left|\Downarrow\right\rangle  & =\left|-,\uparrow\right\rangle -\frac{\lambda_{R}}{2\lambda_{SO}}\left(-i\mathbf{k}_{x}-\mathbf{k}_{y}\right)\left|-,\downarrow\right\rangle +...\label{eq:Low_energy_Kramers}
\end{align}
this being a Kramers doublet, with the Hamiltonian is this basis being
given by \citep{Supplement_2024}:
\begin{equation}
H\left(\mathbf{k}\right)=\left[\left(\frac{\mathbf{k}^{2}}{2m}-\mu\right)-\lambda_{SO}\right]\left[c_{\mathbf{k},\Uparrow}^{\dagger}c_{\mathbf{k},\Uparrow}+c_{\mathbf{k},\Downarrow}^{\dagger}c_{\mathbf{k},\Downarrow}\right]+...\label{eq:Kramers_Hamiltonian}
\end{equation}

\subsubsection{\protect\label{subsec:Adding-pairing}Adding a small pairing}

We consider adding 
\begin{equation}
H_{\Delta}\left(\mathbf{k}\right)=\Delta c_{\mathbf{k}+\uparrow}^{\dagger}c_{-\mathbf{k}+\downarrow}^{\dagger}+\Delta^{*}c_{\mathbf{k}-\uparrow}^{\dagger}c_{-\mathbf{k}-\downarrow}^{\dagger}+h.c.\label{eq:Hamiltonian_pairing}
\end{equation}
to the Hamiltonian in Eq. (\ref{eq:Hamiltonian_matrix-2}) which corresponds
to the pairing piece of the Hamiltonian in Eq. (\ref{eq:Hamiltonian-3-2})..
Then we have that:
\begin{align}
 & T\sum_{\mathbf{k}}H_{\Delta}\left(\mathbf{k}\right)T=\nonumber \\
 & =\sum_{\mathbf{k}}\left[-\Delta^{*}c_{-\mathbf{k}-\downarrow}^{\dagger}c_{\mathbf{k}-\uparrow}^{\dagger}-\Delta c_{-\mathbf{k}+\downarrow}^{\dagger}c_{\mathbf{k}-\uparrow}+h.c.\right]\nonumber \\
 & =\sum_{\mathbf{k}}H_{\Delta}\left(\mathbf{k}\right),\label{eq:TR_invariance}
\end{align}
which means the pairing term in Eq. (\ref{eq:Hamiltonian_pairing})
is time reversal invariant. Now we consider the case where $\Delta$
is the smallest energy scale to the Hamiltonian in Eq. (\ref{eq:Hamiltonian_matrix-2}).
As such it is sufficient to do zeroth order perturbation theory in
$\Delta$. Now we write \citep{Supplement_2024}:
\begin{align}
c_{\mathbf{k}+\uparrow}^{\dagger} & =-\frac{\lambda_{R}}{2\lambda_{SO}}\left(-i\mathbf{k}_{x}-\mathbf{k}_{y}\right)c_{\mathbf{k}\Uparrow}^{\dagger}+...\nonumber \\
c_{\mathbf{k}+\downarrow}^{\dagger} & =c_{\mathbf{k}\Uparrow}^{\dagger}+...\nonumber \\
c_{\mathbf{k}-\uparrow}^{\dagger} & =c_{\mathbf{k}\Downarrow}^{\dagger}+...\nonumber \\
c_{\mathbf{k}-\downarrow}^{\dagger} & =-\frac{\lambda_{R}}{2\lambda_{SO}}\left(+i\mathbf{k}_{x}-\mathbf{k}_{y}\right)c_{\mathbf{k}\Downarrow}^{\dagger}+...\label{eq:Basis_change}
\end{align}
As such computing matrix elements of the Hamiltonian in Eq. (\ref{eq:Hamiltonian_pairing})
we see that within the low energy subspace:
\begin{align}
H_{\Delta}\left(\mathbf{k}\right)= & \frac{\lambda_{R}\Delta}{2\lambda_{SO}}\left(i\mathbf{k}_{x}-\mathbf{k}_{y}\right)c_{\mathbf{k}\Uparrow}^{\dagger}c_{-\mathbf{k}\Uparrow}^{\dagger}+h.c.\nonumber \\
 & -\frac{\lambda_{R}\Delta^{*}}{2\lambda_{SO}}\left(-i\mathbf{k}_{x}-\mathbf{k}_{y}\right)c_{\mathbf{k}\Downarrow}^{\dagger}c_{-\mathbf{k}\Downarrow}^{\dagger}+h.c.\label{eq:Pairing_low_energy}
\end{align}
As such the total low energy effective Hamiltonian is given by: 
\begin{align}
H\left(\mathbf{k}\right)= & \left[\left(\frac{\mathbf{k}^{2}}{2m}-\mu\right)-\lambda_{SO}\right]\left[c_{\mathbf{k},\Uparrow}^{\dagger}c_{\mathbf{k},\Uparrow}+c_{\mathbf{k},\Downarrow}^{\dagger}c_{\mathbf{k},\Downarrow}\right]\nonumber \\
 & +\left[\frac{\lambda_{R}\Delta}{2\lambda_{SO}}\left(-i\mathbf{k}_{x}-\mathbf{k}_{y}\right)c_{\mathbf{k}\Uparrow}^{\dagger}c_{-\mathbf{k}\Uparrow}^{\dagger}\right.\nonumber \\
 & \;\left.-\frac{\lambda_{R}\Delta^{*}}{2\lambda_{SO}}\left(i\mathbf{k}_{x}-\mathbf{k}_{y}\right)c_{\mathbf{k}\Downarrow}^{\dagger}c_{-\mathbf{k}\Downarrow}^{\dagger}+h.c.\right]\label{eq:Low_energy_effective_2D}
\end{align}
Which is a 2D TR invariant Hamiltonian with pseudospin degree of freedom:
$\Uparrow,\,\Downarrow$.

\subsection{\protect\label{subsec:Stability-analysis}Stability analysis}

It is known that superconductors whose effective theory is given by
Eq. (\ref{eq:Low_energy_effective_2D}) have a pair of Majorana Fermions
inside vortex cores \citep{Shen_2017,Bernevig_2013}, denoted by $\gamma_{1}$
and $\gamma_{2}$. We now show that these modes are stable until a
gap closing transition. Indeed from Brillouin-Wigner perturbation
theory we know that the zero energy manifold for a vortex has an effective
Hamiltonian which converges until there is a gap closing phase transition
\citep{Hubac_2010}. By requiring the Hamiltonian be Hermitian and
using fermion parity conservation we must have that the effective
Hamiltonian for the zero energy subspace is given by: 
\begin{equation}
H_{eff}=i\Gamma\gamma_{1}\gamma_{2}\label{eq:Effective_Hamiltonian}
\end{equation}
For some $\Gamma$. If we further assume the perturbation preserves
time reversal $T$ we must have that: 
\begin{equation}
Ti\Gamma\gamma_{1}\gamma_{2}T^{-1}=-i\Gamma\gamma_{1}\gamma_{2}=i\Gamma\gamma_{1}\gamma_{2}\Rightarrow H_{eff}=0\label{eq:zero_effective_Hamiltonian}
\end{equation}
so the Majorana modes are stable (cannot open a gap) and hence the
band topology is stable under small TR preserving perturbations as
well.

\section{\protect\label{sec:1-D-case-(will}1-D case: Doubling the Oreg-Refael-von
Oppen proposal}

\subsection{\protect\label{subsec:Global-analysis-1}Global analysis}

We consider a total (doubled) Hamiltonian given by: 
\begin{align}
H\left(k\right) & =\mu_{z}\left[\left(\frac{k^{2}}{2m}-\mu\right)+\lambda_{SO}\sigma_{z}\tau_{z}+\lambda_{R}k\sigma_{x}\right]\nonumber \\
 & +\Delta_{s}\mu_{x}\label{eq:Hamiltonian-3-2-1}
\end{align}
We note that $\left[H\left(k\right),\tau_{z}\right]=0$ so the two
bands decouple and the Hamiltonian is doubled. We now use the basis:
\begin{align}
\Psi_{k}^{\tau_{z}}= & \left(c_{k+\uparrow},c_{k+\downarrow},c_{-k+\downarrow}^{\dagger},-c_{-k+\uparrow}^{\dagger}\right.\nonumber \\
 & \left.c_{k-\uparrow},c_{k-\downarrow},c_{-k-\downarrow}^{\dagger},-c_{-k-\uparrow}^{\dagger}\right)^{T}\label{eq:Basis_decoupled_1D}
\end{align}
So that the Hamiltonian is explicitly doubled:

\begin{equation}
H\left(k\right)=\left(\begin{array}{cc}
H_{+}\left(k\right) & 0\\
0 & H_{-}\left(k\right)
\end{array}\right)\label{eq:Block_diagonal-1}
\end{equation}
with 
\begin{equation}
H_{\pm}\left(k\right)=\mu_{z}\left[\left(\frac{k^{2}}{2m}-\mu\right)\pm\lambda_{SO}\sigma_{z}+\lambda_{R}k\sigma_{x}\right]+\Delta_{s}\mu_{x}\label{eq:Two_copies-1}
\end{equation}
being two time reversed copies of the Oreg-Refael-von Oppen Hamiltonians
\citep{Oreg_2010,Luchtyn_2010}. Therefore the phase boundaries are
identical to those in that proposal with the critical boundary be
given by the gap closing transition \citep{Oreg_2010} given by Eq.
(\ref{eq:Critical}) above.

\subsection{\protect\label{subsec:Effective-low-energy-1}Effective low energy
theory}

\subsubsection{\protect\label{subsec:Hamiltonian-no-pairing-1}Hamiltonian no pairing}

We now introduce the spinor $\Psi_{k}=\left(c_{k+\uparrow},c_{k+\downarrow},c_{k-\uparrow},c_{k-\downarrow}\right)^{T}$
then we consider the Hamiltonian given by: 
\begin{equation}
H\left(k\right)=\Psi_{k}^{\dagger}\mathcal{H}\left(k\right)\Psi_{k}\label{eq:Hamiltonian_matrix-2-1}
\end{equation}
with:
\begin{equation}
\mathcal{H}\left(k\right)=\left(\frac{k^{2}}{2m}-\mu\right)+\lambda_{SO}\tau_{z}\sigma_{z}+\lambda_{R}k\sigma_{x}\label{eq:4_by_4-1}
\end{equation}
It is straightforward to check that $\left[\mathcal{H}\left(k\right),\tau_{z}\right]=0$
and that the Hamiltonian in Eq. (\ref{eq:4_by_4-1}) is time reversal
invariant and corresponds to the non-pairing piece of the Hamiltonian
in Eq. (\ref{eq:Hamiltonian-3-2-1}). Now we will assume that $\lambda_{SO}\gg\lambda_{R}$
so that the low energy Hamiltonian is given by \citep{Supplement_2024}:
\begin{align}
\left|\Uparrow\right\rangle  & =\left|+,\downarrow\right\rangle -\frac{\lambda_{R}}{2\lambda_{SO}}k\left|+,\uparrow\right\rangle +...\nonumber \\
\left|\Downarrow\right\rangle  & =\left|-,\uparrow\right\rangle -\frac{\lambda_{R}}{2\lambda_{SO}}k\left|-,\downarrow\right\rangle +...\label{eq:Low_energy_Kramers-1}
\end{align}
being a Kramers doublet, with 
\begin{equation}
H\left(\mathbf{k}\right)=\left[\left(\frac{k^{2}}{2m}-\mu\right)-\lambda_{SO}\right]\left[c_{k,\Uparrow}^{\dagger}c_{k,\Uparrow}+c_{k,\Downarrow}^{\dagger}c_{k,\Downarrow}\right]+...\label{eq:Kramers_Hamiltonian-1}
\end{equation}
where we have focused just on the low energy subspace.

\subsubsection{\protect\label{subsec:Adding-pairing-1}Adding pairing}

We consider adding 
\begin{equation}
H_{\Delta}\left(k\right)=\Delta c_{k+\uparrow}^{\dagger}c_{-k+\downarrow}^{\dagger}+\Delta^{*}c_{k-\uparrow}^{\dagger}c_{-k-\downarrow}^{\dagger}+h.c.\label{eq:Hamiltonian_pairing-1}
\end{equation}
to the Hamiltonian in Eq. (\ref{eq:Hamiltonian_matrix-2-1}) which
corresponds to the pairing piece of the Hamiltonian in Eq. (\ref{eq:Hamiltonian-3-2-1})
and is TR invariant see Eq. (\ref{eq:TR_invariance}). Now we write
\citep{Supplement_2024}:
\begin{align}
c_{k+\uparrow}^{\dagger} & =-\frac{\lambda_{R}}{2\lambda_{SO}}kc_{k\Uparrow}^{\dagger}+...\nonumber \\
c_{k+\downarrow}^{\dagger} & =c_{k\Uparrow}^{\dagger}+...\nonumber \\
c_{k-\uparrow}^{\dagger} & =c_{k\Downarrow}^{\dagger}+...\nonumber \\
c_{k-\downarrow}^{\dagger} & =-\frac{\lambda_{R}}{2\lambda_{SO}}kc_{k\Downarrow}^{\dagger}+...\label{eq:Basis_change-1}
\end{align}
As such we have that within the low energy subspace \citep{Supplement_2024}:
\begin{equation}
H_{\Delta}\left(k\right)=\frac{\lambda_{R}\Delta}{2\lambda_{SO}}kc_{k\Uparrow}^{\dagger}c_{-k\Uparrow}^{\dagger}-\frac{\lambda_{R}\Delta^{*}}{2\lambda_{SO}}kc_{k\Downarrow}^{\dagger}c_{-k\Downarrow}^{\dagger}+h.c.\label{eq:Pairing_low_energy-1}
\end{equation}
As such the total low energy effective Hamiltonian is given by \citep{Supplement_2024}:
\begin{align}
H\left(k\right) & =\left[\left(\frac{k^{2}}{2m}-\mu\right)-\lambda_{SO}\right]\left[c_{k,\Uparrow}^{\dagger}c_{k,\Uparrow}+c_{k,\Downarrow}^{\dagger}c_{k,\Downarrow}\right]\nonumber \\
 & +\left[\frac{\lambda_{R}\Delta}{2\lambda_{SO}}kc_{k\Uparrow}^{\dagger}c_{-k\Uparrow}^{\dagger}-\frac{\lambda_{R}\Delta^{*}}{2\lambda_{SO}}kc_{k\Downarrow}^{\dagger}c_{-k\Downarrow}^{\dagger}+h.c.\right]\label{eq:1D_TR_invariant}
\end{align}
Which is a p-wave 1D time reversal invariant Hamiltonian with pseudospin
degree of freedom.

\subsection{\protect\label{subsec:Stability-theory}Stability analysis}

The stability analysis is verbatim that of Section \ref{subsec:Stability-analysis}.

\section{\protect\label{sec:Conclusions}Conclusions \& outlook}

In this work we have doubled in a TR preserving way the Sau-Luchtin-Tewari-Sarma
\citep{Sau_2009} and Oreg-Refael-von Oppen \citep{Oreg_2010,Luchtyn_2010}
proposals much like the Kane-Mele proposal (spin Hall insulator) \citep{Kane_2005}
doubles the Haldane model \citep{Haldane_1988} in a TR preserving
way. We have shown that the low energy effective action for these
models in 1D and 2D are the time reversal invariant single band p-wave
superconductors with pseudospin degree of freedom. Similarly the effective
action for the Sau-Luchtin-Tewari-Sarma \citep{Sau_2009} and Oreg-Refael-von
Oppen \citep{Oreg_2010,Luchtyn_2010} are the single band $p+ip$
and single band 1D p-wave superconductors with spin degree of freedom.
In the supplement \citep{Supplement_2024} we have presented physical
realization of the system within the cold atoms setup \citep{Haim_2019,Haim_2016}.
In future works it would be of interest to study real solid state
systems with spin orbit coupling and proximity induces s-wave superconductivity
for a completely realistic realization of the models presented in
this work in solid state heterostructures which would open this proposal
to many applications in quantum computing \citep{Nielsen_2011}.

\textbf{Acknowledgements:} The author would like to thank Chris Laumann
and Claudio Chamon for useful discussions.

\appendix

\part*{Supplementary Online Information}

\section{\protect\label{sec:Background-(lightening-review)}Background (lightening
review)}

\subsection{\protect\label{subsec:Topological-supercondcutivity}Topological
superconductivity}

\subsubsection{\protect\label{subsec:Time-reversal-symmetry}Time reversal symmetry
breaking p-wave 2D superconductors}

The simplest time reversal symmetry breaking p-wave 2D superconductors
have the following Hamiltonian: 
\begin{align}
H\left(\mathbf{k}\right) & =\left(\frac{\mathbf{k}^{2}}{2m}-\mu\right)c_{\mathbf{k}}^{\dagger}c_{\mathbf{k}}\nonumber \\
 & +\Delta\left(\mathbf{k}_{x}\pm i\mathbf{k}_{y}\right)c_{\mathbf{k}}^{\dagger}c_{-\mathbf{k}}^{\dagger}+h.c.\label{eq:chiral_p_wave}
\end{align}
Where $\pm$ corresponds to $p_{x}+ip_{y}$ and $p_{x}-ip_{y}$ Hamiltonians.
It is known to have Majorana fermions in the vortices of its order
parameter \citep{Shen_2017,Bernevig_2013}.

\subsubsection{\protect\label{subsec:Time-reversal-symmetry-1}Time reversal symmetry
invariant p-wave 2D Hamiltonians}

The simplest time reversal symmetry preserving p-wave 2D superconductors
have the following Hamiltonian \citep{Haim_2019}: 
\begin{align}
H\left(\mathbf{k}\right)= & \sum_{\sigma=\pm}\left(\frac{\mathbf{k}^{2}}{2m}-\mu\right)c_{\mathbf{k}\sigma}^{\dagger}c_{\mathbf{k}\sigma}\nonumber \\
 & +\Delta\left(\mathbf{k}_{x}\pm i\mathbf{k}_{y}\right)c_{\mathbf{k}\uparrow}^{\dagger}c_{-\mathbf{k}\uparrow}^{\dagger}+h.c.\nonumber \\
 & -\Delta^{*}\left(\mathbf{k}_{x}\mp i\mathbf{k}_{y}\right)c_{\mathbf{k}\downarrow}^{\dagger}c_{-\mathbf{k}\downarrow}^{\dagger}+h.c.\label{eq:Time_reversal_p_wave}
\end{align}
Which is just two time reversal invariant copies of the Hamiltonian
in Eq. (\ref{eq:chiral_p_wave}).

\subsubsection{\protect\label{subsec:Time-reversal-symmetry-3}Time reversal symmetry
breaking p-wave 1D superconductors}

The simplest time reversal symmetry breaking p-wave 1D superconductors
have the following Hamiltonian: 
\begin{equation}
H\left(k\right)=\left(\frac{k^{2}}{2m}-\mu\right)c_{k}^{\dagger}c_{k}+\left[\Delta kc_{k}^{\dagger}c_{-k}^{\dagger}+h.c.\right]\label{eq:Hamiltonian-2}
\end{equation}
It is known to have Majorana fermions at the ends of 1D wires \citep{Shen_2017,Bernevig_2013}.

\subsubsection{\protect\label{subsec:Time-reversal-symmetry-1-1}Time reversal symmetry
invariant p-wave 1D Hamiltonians}

The simplest time reversal symmetry preserving p-wave 2D superconductors
have the following Hamiltonian \citep{Haim_2019}:
\begin{align}
H\left(k\right)= & \sum_{\sigma=\pm}\left(\frac{k^{2}}{2m}-\mu\right)c_{\mathbf{k}\sigma}^{\dagger}c_{\mathbf{k}\sigma}\nonumber \\
 & +\left[\Delta kc_{k\uparrow}^{\dagger}c_{-k\uparrow}^{\dagger}-\Delta^{*}kc_{k\downarrow}^{\dagger}c_{-k\downarrow}^{\dagger}+h.c.\right]\label{eq:1D_TR_p_wave}
\end{align}
 Which is just two time reversal invariant copies of the Hamiltonian
in Eq. (\ref{eq:Hamiltonian-2}).

\subsection{\protect\label{subsec:Quantum-mechanics}Quantum mechanics}

\subsubsection{\protect\label{subsec:First-order-perturbation}First order perturbation
theory}

We note that if 
\begin{equation}
H_{0}\left|E_{0}\right\rangle =E_{0}\left|E_{0}\right\rangle \label{eq:Spectrum}
\end{equation}
and $H=H_{0}+V$ with $V\ll H_{0}$ then:
\begin{align}
H\left|E\right\rangle  & =E\left|E\right\rangle \nonumber \\
E & =E_{0}+\left\langle E_{0}\right|V\left|E_{0}\right\rangle +.....\nonumber \\
\left|E\right\rangle  & =\left|E_{0}\right\rangle -\sum_{E_{0}'\neq E_{0}}\frac{\left\langle E_{0}'\right|V\left|E_{0}\right\rangle }{E_{0}'-E_{0}}\left|E_{0}'\right\rangle +....\label{eq:Rayleigh_schordinger}
\end{align}

\subsubsection{\protect\label{subsec:Change-of-basis}Change of basis}

Suppose you have an orthonormal basis $\left\{ \left|U_{i}\right\rangle _{i=1,..N}\right\} $
and a second orthonormal basis $\left\{ \left|V_{i}\right\rangle _{i=1,...N}\right\} $
then we have that: 
\begin{equation}
\left|U_{j}\right\rangle =\sum_{i=1}^{N}\left|V_{i}\right\rangle \left\langle V_{i}\mid U_{j}\right\rangle \label{eq:Change_basis}
\end{equation}
Now suppose there is some relevant Hamiltonian such that $\left|V_{i}\right\rangle _{i=1,..M}$
is much lower in energy then $\left|V_{i}\right\rangle _{i=M+1,...N}$
then in many cases we may write that: 
\begin{equation}
\left|U_{j}\right\rangle =\sum_{i=1}^{M}\left|V_{i}\right\rangle \left\langle V_{i}\mid U_{j}\right\rangle +....\label{eq:Basis_truncation}
\end{equation}

\section{\protect\label{sec:Proximity-induced-superconductiv}Proximity induced
superconductivity}

\subsection{\protect\label{subsec:Sau-Luchtin-Tewari-Sarma-proposa}Sau-Luchtin-Tewari-Sarma
proposal \citep{Sau_2009} (simplified treatment)}

Consider the following Hamiltonian: 
\begin{equation}
H\left(\mathbf{k}\right)=\left(c_{\mathbf{k}\uparrow}^{\dagger},c_{\mathbf{k}\downarrow}^{\dagger}\right)h\left(\mathbf{k}\right)\left(\begin{array}{c}
c_{\mathbf{k\uparrow}}\\
c_{\mathbf{k}\downarrow}
\end{array}\right)\label{eq:Hamiltonian_matrix}
\end{equation}
with 
\begin{align}
h\left(\mathbf{k}\right) & =\left(\frac{\mathbf{k}^{2}}{2m}-\mu\right)+\lambda_{R}\left(\mathbf{k}_{x}\sigma_{y}-\mathbf{k}_{y}\sigma_{x}\right)+B\sigma_{z}\nonumber \\
h_{0}\left(\mathbf{k}\right) & =\left(\frac{\mathbf{k}^{2}}{2m}-\mu\right)+B\sigma_{z}\nonumber \\
V\left(\mathbf{k}\right) & =\lambda_{R}\left(\mathbf{k}_{x}\sigma_{y}-\mathbf{k}_{y}\sigma_{x}\right)\label{eq:Pertubation}
\end{align}
Then using Eq. (\ref{eq:Rayleigh_schordinger}) we have that 
\begin{align}
\left|\Omega\right\rangle  & =\left|\downarrow\right\rangle -\frac{\lambda_{R}}{2B}\left(i\mathbf{k}_{x}-\mathbf{k}_{y}\right)\left|\uparrow\right\rangle +...\nonumber \\
E_{\Omega} & =\left(\frac{\mathbf{k}^{2}}{2m}-\mu\right)-B+....\label{eq:Ground_state}
\end{align}
Where we have assumed that $B\gg\lambda_{R}$. Now we write: 
\begin{align}
c_{\mathbf{k}\downarrow}^{\dagger} & =c_{\mathbf{k},\Omega}^{\dagger}+...\nonumber \\
c_{\mathbf{k}\uparrow}^{\dagger} & =-\frac{\lambda_{R}}{2B}\left(-i\mathbf{k}_{x}-\mathbf{k}_{y}\right)c_{\mathbf{k},\Omega}^{\dagger}+...\label{eq:Creation_operators}
\end{align}
Now we add s-wave paring to the Hamiltonian in Eq. (\ref{eq:Hamiltonian_matrix})
with the new Hamiltonian being given by: 

\begin{equation}
H\left(\mathbf{k}\right)=\left(c_{\mathbf{k}\uparrow}^{\dagger},c_{\mathbf{k}\downarrow}^{\dagger}\right)h\left(\mathbf{k}\right)\left(\begin{array}{c}
c_{\mathbf{k\uparrow}}\\
c_{\mathbf{k}\downarrow}
\end{array}\right)+\left[\Delta c_{\mathbf{k}\uparrow}^{\dagger}c_{-\mathbf{k}\downarrow}^{\dagger}+h.c.\right]\label{eq:Hamiltonian_matrix-1}
\end{equation}
Now within the low energy subspace spanned by $\left|\Omega\right\rangle $
we have that: 
\begin{align}
H_{\Omega}\left(\mathbf{k}\right)= & \left[\left(\frac{\mathbf{k}^{2}}{2m}-\mu\right)-B\right]c_{\mathbf{k},\Omega}^{\dagger}c_{\mathbf{k},\Omega}\nonumber \\
 & +\left[\frac{\lambda_{R}\Delta}{2B}c_{\mathbf{k}\Omega}^{\dagger}c_{-\mathbf{k}\Omega}^{\dagger}\left(\mathbf{k}_{y}+i\mathbf{k}_{x}\right)+h.c.\right]\label{eq:Low_energy_Sau}
\end{align}
or a 2D p-wave time reversal breaking Hamiltonian.

\subsection{\protect\label{subsec:Sau-Luchtin-Tewari-Sarma-proposa-1} Oreg-Refael-von
Oppen proposal \citep{Oreg_2010,Luchtyn_2010} (simplified treatment)}

Consider the following Hamiltonian: 
\begin{equation}
H\left(k\right)=\left(c_{k\uparrow}^{\dagger},c_{k\downarrow}^{\dagger}\right)h\left(k\right)\left(\begin{array}{c}
c_{k\uparrow}\\
c_{k\downarrow}
\end{array}\right)\label{eq:Hamiltonian_matrix-3}
\end{equation}
with 
\begin{align}
h\left(k\right) & =\left(\frac{k^{2}}{2m}-\mu\right)+\lambda_{R}k\sigma_{x}+B\sigma_{z}\nonumber \\
h_{0}\left(k\right) & =\left(\frac{k^{2}}{2m}-\mu\right)+B\sigma_{z}\nonumber \\
V\left(k\right) & =\lambda_{R}k\sigma_{x}\label{eq:Pertubation-1}
\end{align}
Then using Eq. (\ref{eq:Rayleigh_schordinger}) we have that 
\begin{align}
\left|\Omega\right\rangle  & =\left|\downarrow\right\rangle -\frac{\lambda_{R}}{2B}k\left|\uparrow\right\rangle +...\nonumber \\
E_{\Omega} & =\left(\frac{k^{2}}{2m}-\mu\right)-B+....\label{eq:Ground_state-1}
\end{align}
Where we have assumed that $B\gg\lambda_{R}$. Now we write: 
\begin{align}
c_{k\downarrow}^{\dagger} & =c_{k\Omega}^{\dagger}+...\nonumber \\
c_{k\uparrow}^{\dagger} & =-\frac{\lambda_{R}}{2B}kc_{k,\Omega}^{\dagger}+...\label{eq:Creation_operators-1}
\end{align}
Now we add s-wave paring to the Hamiltonian in Eq. (\ref{eq:Hamiltonian_matrix-3})
with the new Hamiltonian being given by: 

\begin{equation}
H\left(k\right)=\left(c_{k\uparrow}^{\dagger},c_{k\downarrow}^{\dagger}\right)h\left(k\right)\left(\begin{array}{c}
c_{k\uparrow}\\
c_{k\downarrow}
\end{array}\right)+\left[\Delta c_{k\uparrow}^{\dagger}c_{-k\downarrow}^{\dagger}+h.c.\right]\label{eq:Hamiltonian_matrix-1-1}
\end{equation}
Now within the low energy subspace spanned by $\left|\Omega\right\rangle $
we have that: 
\begin{align}
H_{\Omega}\left(k\right)= & \left[\left(\frac{k^{2}}{2m}-\mu\right)-B\right]c_{k,\Omega}^{\dagger}c_{k,\Omega}\nonumber \\
 & +\frac{\lambda_{R}\Delta}{2B}c_{k\Omega}^{\dagger}c_{-k\Omega}^{\dagger}k+h.c.\label{eq:Low_energy_oreg}
\end{align}
or a 1D p-wave time reversal breaking Hamiltonian.

\section{\protect\label{sec:Cold-atoms-realizations}Cold atoms realizations
of the proposals in the main text}

\begin{figure}
\begin{centering}
\includegraphics[width=1\columnwidth]{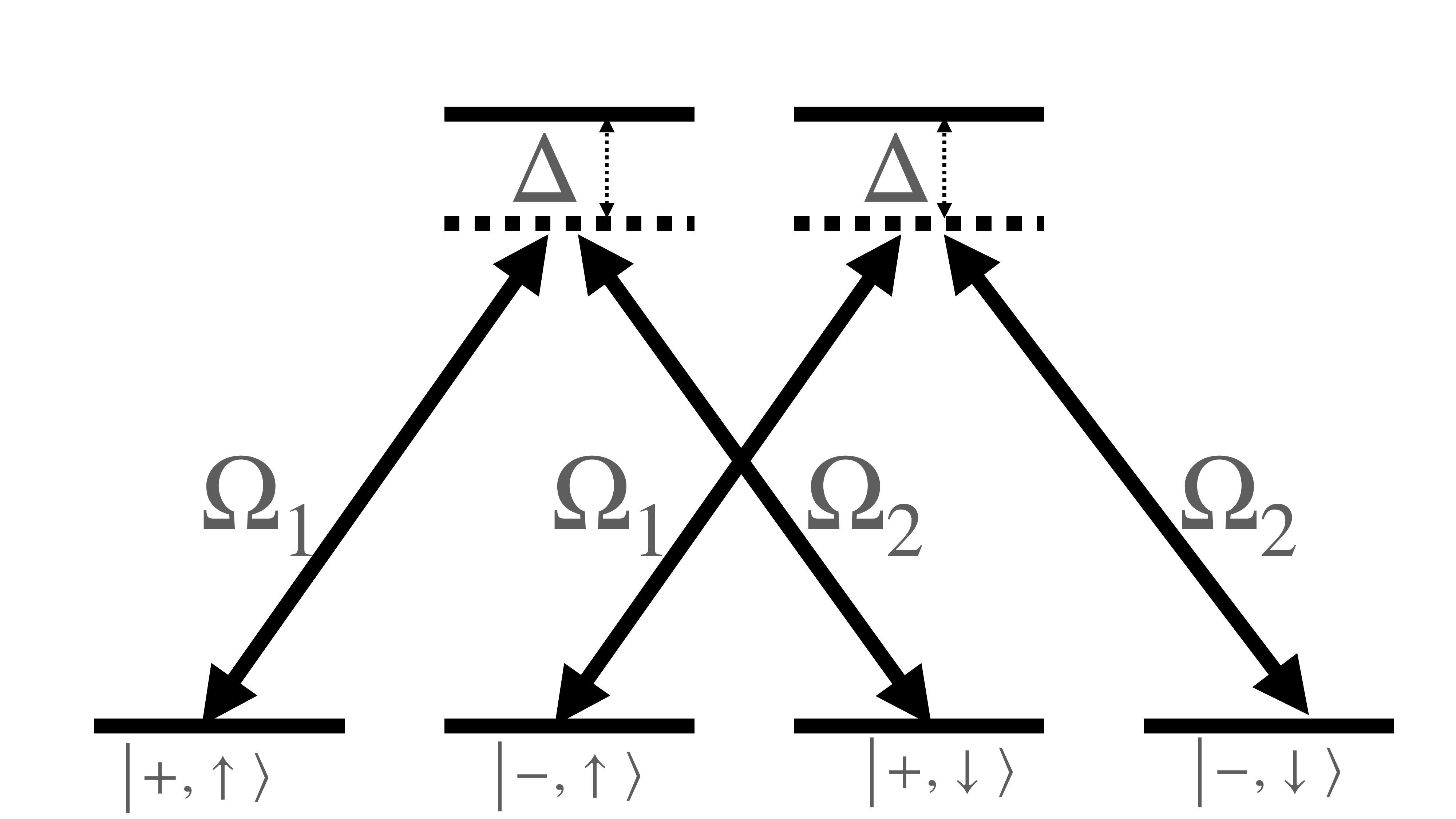}
\par\end{centering}
\caption{\protect\label{fig-lasers} Thelase setup for the could atom realization.}
\end{figure}

We would like to emulate the Hamiltonian given in Ref. \citep{Keselman_2013}
which is similar to ours but has different time reversal properties
using cold atoms. The Hamiltonian we would like to emulate is given
by: 
\begin{align}
H\left(k\right) & =\mu_{z}\left[\left(\frac{k^{2}}{2m}-\mu\right)+\lambda_{SO}\sigma_{x}+\lambda_{R}k\sigma_{z}\tau_{z}\right]\nonumber \\
 & +\Delta_{s}\mu_{x}\tau_{z}\label{eq:Hamiltonian-3-2-1-1}
\end{align}
We will closely be following \citep{Jiang_2011}. Now we consider
the setup with four lasers see Fig. (\ref{fig-lasers}). We have that
the Hamiltonian is given by:
\begin{align}
H & =\sum_{k}\left(\frac{k^{2}}{2m}-V\right)n_{k}+\nonumber \\
 & B\sum_{k}\left(c_{k+p+\uparrow}^{\dagger}c_{k-p+\downarrow}+c_{k+p-\uparrow}^{\dagger}c_{k-p-\downarrow}+h.c.\right)\label{eq:Coupling}
\end{align}
Where 
\begin{align}
n_{k} & =c_{k+\uparrow}^{\dagger}c_{k+\uparrow}+c_{k+\downarrow}^{\dagger}c_{k+\downarrow}+c_{k-\uparrow}^{\dagger}c_{k-\uparrow}+c_{k-\downarrow}^{\dagger}c_{k-\downarrow}\nonumber \\
B & =\frac{\Omega_{1}\Omega_{2}^{*}}{\Delta}\label{eq:Field}
\end{align}
and $p$ is the momentum of the lasers. We will now assume two species
of molecules with the Hamiltonian: 
\begin{align}
H_{Fes} & =g\int dxb_{+}\left(x\right)c_{+\uparrow}^{\dagger}\left(x\right)c_{+\downarrow}^{\dagger}\left(x\right)+h.c.\nonumber \\
 & +g\int dxb_{-}\left(x\right)c_{-\uparrow}^{\dagger}\left(x\right)c_{-\downarrow}^{\dagger}\left(x\right)+h.c.\label{eq:Fesbach}
\end{align}
Now we assume that due to molecule condensation and an order from
disorder effect 
\begin{equation}
g\left\langle b_{+}\left(x\right)\right\rangle =-g\left\langle b_{-}\left(x\right)\right\rangle =\Xi\label{eq:Condensation}
\end{equation}
This means that:
\begin{align}
H & =\sum_{k}\left(\frac{k^{2}}{2m}-V\right)n_{k}\nonumber \\
 & +\sum_{k}B\left(c_{k+p+\uparrow}^{\dagger}c_{k-p+\downarrow}+c_{k+p-\uparrow}^{\dagger}c_{k-p-\downarrow}+h.c.\right)\nonumber \\
 & +\sum_{k}\Xi\left(c_{k+\uparrow}^{\dagger}c_{-k+\downarrow}^{\dagger}-c_{k-\uparrow}^{\dagger}c_{-k-\downarrow}^{\dagger}+h.c.\right)\label{eq:Pairing_Hamiltonian}
\end{align}
 We now perform the transform 
\begin{equation}
\exp\left(ip\int x\left(n_{+\uparrow}\left(x\right)-n_{+\downarrow}\left(x\right)-n_{-\uparrow}\left(x\right)+n_{-\downarrow}\left(x\right)\right)\right)\label{eq:Transform-4}
\end{equation}
This transforms the Hamiltonian to the form: 
\begin{align}
H & =\sum_{k}\sum_{\sigma_{z}=\pm;\tau_{z}=\pm}\left(\frac{\left(k-p\sigma_{z}\tau_{z}\right)^{2}}{2m}-V\right)n_{k\sigma_{z}\tau_{z}}\nonumber \\
 & +\sum_{k}B\left(c_{k+\uparrow}^{\dagger}c_{k+\downarrow}+c_{k-\uparrow}^{\dagger}c_{k-\downarrow}+h.c.\right)\nonumber \\
 & +\sum_{k}\Xi\left(c_{k-p+\uparrow}^{\dagger}c_{-k+p+\downarrow}^{\dagger}-c_{k-p-\uparrow}^{\dagger}c_{-k+p-\downarrow}^{\dagger}+h.c.\right)\nonumber \\
 & =\sum_{k}\left(\frac{k^{2}}{2m}-V+\frac{p^{2}}{2m}\right)n_{k}\nonumber \\
 & -\sum_{k}\sum_{\sigma_{z}=\pm;\tau_{z}=\pm}\left(\frac{kp\sigma_{z}\tau_{z}}{m}\right)n_{k\sigma_{z}\tau_{z}}\nonumber \\
 & +\sum_{k}B\left(c_{k+\uparrow}^{\dagger}c_{k+\downarrow}+c_{k-\uparrow}^{\dagger}c_{k-\downarrow}+h.c.\right)\nonumber \\
 & +\sum_{k}\Xi\left(c_{k-p+\uparrow}^{\dagger}c_{-k+p+\downarrow}^{\dagger}-c_{k-p-\uparrow}^{\dagger}c_{-k+p-\downarrow}^{\dagger}+h.c.\right)\label{eq:Final_Hamiltonian}
\end{align}

\section{\protect\label{sec:Comparison-with-Ref.}Comparison with Ref. \citep{Haim_2019}}

In Appendix A of Ref. \citep{Haim_2019} the authors present results
that the constructions presented in the main text are impossible.
We make no quarrel with the math presented in Ref. \citep{Haim_2019}
however their main assumption in Appendix A is that the paring Hamiltonian
multiplied by the time reversal matrix is positive semidefinite (in
which case, to summarize their results, the order parameter cannot
wind in $\mathbf{k}$ space, see e.g. Eq. (\ref{eq:chiral_p_wave})
so the superconductor is trivial) is too restrictive. However if we
consider the form of Eq. (\ref{eq:Time_reversal_transform}) we see
their main assumption about positive semi definiteness does not apply
to our setup. Indeed we see that: 
\begin{align}
 & T\Delta_{\mathbf{k}}c_{\mathbf{k}+\uparrow}^{\dagger}\rightarrow-\Delta Tc_{-\mathbf{k}+\downarrow}^{\dagger}\rightarrow-\Delta c_{\mathbf{k}-\uparrow}^{\dagger},\nonumber \\
 & T\Delta_{\mathbf{k}}c_{\mathbf{k}+\downarrow}^{\dagger}\rightarrow\Delta Tc_{-\mathbf{k}+\downarrow}^{\dagger}\rightarrow-\Delta c_{\mathbf{k}-\downarrow}^{\dagger},\nonumber \\
 & T\Delta_{\mathbf{k}}c_{\mathbf{k}-\uparrow}^{\dagger}\rightarrow-\Delta^{*}Tc_{-\mathbf{k}-\downarrow}^{\dagger}\rightarrow-\Delta^{*}c_{\mathbf{k}+\uparrow}^{\dagger},\nonumber \\
 & T\Delta_{\mathbf{k}}c_{\mathbf{k}-\downarrow}^{\dagger}\rightarrow\Delta^{*}Tc_{-\mathbf{k}-\downarrow}^{\dagger}\rightarrow-\Delta^{*}c_{\mathbf{k}+\downarrow}^{\dagger}\label{eq:Transform}
\end{align}
Where $\Delta_{\mathbf{k}}$ is the restriction of the pairing matrix
to the wavevector $\mathbf{k}$. As such we have that: 
\begin{equation}
T\Delta_{\mathbf{k}}=-\left(\begin{array}{cccc}
0 & 0 & \Delta^{*} & 0\\
0 & 0 & 0 & \Delta^{*}\\
\Delta & 0 & 0 & 0\\
0 & \Delta & 0 & 0
\end{array}\right)\label{eq:Mixed}
\end{equation}
is not positive or negative semidefinite as it has eigenvalues $\pm\left|\Delta\right|$,
as such the main assumption of Appendix A of Ref. \citep{Haim_2019}
does not apply to our setup.

\end{document}